 \documentclass[final,5p,times,twocolumn]{elsarticle}

\usepackage{xcolor}

\usepackage[switch]{lineno}

\usepackage{amsmath}
\usepackage{graphicx}
\usepackage{subfigure}
\usepackage{multicol} 

\usepackage{amssymb}
\usepackage{bm}
\usepackage{algorithm}
\usepackage{algorithmic}
\usepackage{amsthm}
\usepackage{amssymb}
\usepackage{amsmath}
\usepackage{graphicx}
\usepackage{subfigure}
\usepackage{booktabs}
\usepackage{hyperref}

\usepackage[]{caption2}



\usepackage{setspace}%

\journal{\textbf{ }}

\begin{document}

\begin{frontmatter}



\title{Towards the Next Generation of Retinal Neuroprosthesis: Visual Computation with Spikes}


\author[pku,pchen]{Zhaofei Yu}
\author[uk]{Jian K. Liu \corref{cor1}}
\cortext[cor1]{Corresponding author. }
\ead{jian.liu@leicester.ac.uk }
\author[pku,pchen]{Shanshan Jia}
\author[pku,pchen]{Yichen Zhang}
\author[pku,pchen]{Yajing Zheng}
\author[pku,pchen]{Yonghong Tian}
\author[pku,pchen]{Tiejun Huang}
\address[pku]{National Engineering Laboratory for Video Technology, School of Electronics Engineering and Computer Science,  Peking University, Beijing, China}
\address[pchen]{Peng Cheng Laboratory, Shenzhen, China}
\address[uk]{ Centre for Systems Neuroscience, Department of Neuroscience, Psychology and Behaviour, University of Leicester, Leicester, UK}

\begin{abstract}
Neuroprosthesis, as one type of precision medicine device, is aiming for manipulating neuronal signals of the brain in a closed-loop fashion, together with receiving stimulus from the environment and controlling some part of our brain/body. In terms of vision, incoming information can be processed by the brain in millisecond interval. The retina computes visual scenes and then sends its output as neuronal spikes to the cortex for further computation. Therefore,  the neuronal signal of interest for retinal neuroprosthesis is spike. Closed-loop computation in neuroprosthesis includes two stages: encoding stimulus to neuronal signal, and decoding it into stimulus. Here we review some of the recent progress about visual computation models that use spikes for analyzing natural scenes, including static images and dynamic movies. We hypothesize that for a better understanding of computational principles in the retina, one needs a hypercircuit view of the retina, in which different functional network motifs revealed in the cortex neuronal network should be taken into consideration for the retina. Different building blocks of the retina, including a diversity of cell types and synaptic connections, either chemical synapses or electrical synapses (gap junctions), make the retina an ideal neuronal network to adapt the computational techniques developed in artificial intelligence for modeling of encoding/decoding visual scenes. Altogether, one needs a systems approach of visual computation with spikes to advance the next generation of retinal neuroprosthesis as an artificial visual system.
\end{abstract}


\end{frontmatter}



\section{Introduction}

The concept of precision medicine has been proposed for a few years. Mostly, it has been referred to customize healthcare to individual patient. Nowadays, the advancements of artificial intelligence techniques, including both hardware and software/algorithm, make the process of healthcare more precise for each individual patient such that the communication between healthcare device/service and patient is specifically designed and justified. 

Neuroprosthesis is such a precise medicine device. As a way of therapy besides traditional pharmacological treatment, it usually has a direct interaction with neuronal activity, in particular, neuronal spikes, for each individual brain~\cite{chapin1999real,taylor2002direct,musallam2004cognitive,moritz2008direct,velliste2008cortical,gilja2012high,hochberg2012reach,collinger2013high,shanechi2017rapid}. It consists of a series of devices that could substitute some part of our body and/or brain, such as motor, sensory and cognitive modality that has been damaged. As the brain is the central hub to control and exchange information used by our motor, sensory and cognitive behavior, the performance of neuroprosthesis has to rely on how to better analyze the neuronal signal used by neuroprosthesis. Therefore, besides the development of neuroprosthesis hardware, better algorithms are the core feature of neuroprosthesis for better performance~\cite{gilja2012high,Nirenberg2012, seeber2016history}. 

Motor neuroprosthesis has a long history with intensive studies, in particular with the recent techniques where cortical neuronal spikes can be well recorded and used to control neuroprosthesis~\cite{gilja2012high}. In term of sensory neuroprosthesis, cochlear implants are severed as the most widely used neuroprosthesis, and have a fair good performance for helping hearing loss, although there are many remaining questions about how to improve its performance in a noisy environment and its effect on neuronal activity of downstream auditory cortex~\cite{seeber2016history, johnson2017representations}. However, in contrast to the intensive computational modeling of cochlear implants~\cite{seeber2016history}, retinal neuroprosthesis is much less well studied, and has a much worse performance for restoring the eyesight, although a few types of retinal neuroprosthesis are being used in clinical trials~\cite{yue2016retinal, ha2016towards}. 

The retina consists of three layers of neurons with photoreceptors, bipolar cells, and ganglion cells together with inhibitory horizontal and amacrine cells in between. Photoreceptors receive the incoming light signal encoded natural environment and transform it into the electrical activity that is modulated by horizontal cells. Then these activities are sent to bipolar cells and amacrine cells to further processing. In the end, all of these signals go to the output side of the retina, where retinal ganglion cells, as the only output neurons, produce a sequence of action potentials or spikes, which are transmitted via the optic nerve to various downstream brain regions. Essentially, all the visual information about our environment, both in space and time, is encoded by these spatiotemporal patterns of spikes from ganglion cells.

Many types of eye diseases are caused by neuronal degeneration of photoreceptors, whereas the outputs of the retina, ganglion cells, remain healthy. One type of therapy would be to develop an advanced retinal prosthesis to directly stimulate ganglion cells with an array of electrodes. Retinal neuroprosthesis also has a relatively long history of research~\cite{shepherd2013visual}. However, much effort is dedicated to material design of retinal neuroprosthesis hardware~\cite{yue2016retinal,ha2016towards, shepherd2013visual, ghezzi2015retinal,tang2018nanowire, goetz2016electronic}. Recently, it has been suggested that employing better neural coding algorithms were able to improve the performance of retinal neuroprosthesis~\cite{Nirenberg2012}, where it was shown that on top of neuroprosthesis, reconstruction of visual scenes can be significantly improved by adding an encoder converting input images into spiking codes used by retinal ganglion cells, then using these codes to drive transducers, such as electrodes, optogenetic stimulators, or other components for vision restoration. 

Therefore, one needs better computational models to advance the performance of retinal neuroprosthesis. Comparing to other neuroprostheses where stimulus signals are relatively simple, retinal neuroprostheses deal with dynamical visual scenes in space and time with higher order correlations.  Low performance is mainly due to a major difficulty that there is no clear understanding of how ganglion cells encode rich visual scenes. Much of our knowledge has been documented through experiments with simple artificial stimuli, such as white noise images, bars, and gratings, etc. It remains unclear how our retina processes complex natural images with its neuronal underpinnings. In recent years, artificial intelligence has seen remarkable progress in analyzing complex visual scenes, including natural images and movies. Thus, now it is possible to develop novel functional artificial intelligence models to study the encoding and decoding of natural scenes by analyzing retinal ganglion cell spiking responses.

In this paper, we review some of the recent progress on this topic. Most of the studies on visual coding can be roughly classified into two streams. The first and traditional stream can be named as feature-based modeling approach, where visual features or filters can be aligned with some biophysical properties, such as receptive field, of the retinal neurons. The second and relative new stream can be named as sampling-based modeling approach, where statistics of visual scenes, such as pixels, are formulated by some probabilistic models. 
We review some of the core ideas emerged from both approaches for analysis of visual scenes with the utility of neural spikes with the aim for the next generation of retinal neuroprosthesis where computational modelling plays an essential role.

The organization of this review is as follows. 

Sec.~\ref{retina} gives an introduction of the biological underpinnings of the retina with a focus on its inner neuronal circuit. We emphasize that the retinal circuit carries out rich computations that are beyond the dynamics of single cells of the retina. 

In Sec.~\ref{computing}, in contrast to the view that the retina is a simple neural network, we hypothesize that the retina is highly complex and comparable to some aspects of the cortex with different network motifs for specialized computations to extract visual feature. In particular, we outline three views of the retinal neuronal circuit as feedforward, recurrent and winner-take-all network structures. For each of these three viewpoints, we provide some evidence and recent results that fit into the proposed framework. 

In Sec.~\ref{coding}, feature-based modeling approach is discussed, where the models of encoding and decoding visual scenes based on feature extraction by the retina are reviewed. For encoding, we first summarize biophysical models that directly analyze and fit neuronal spikes to obtain some neuronal properties such as receptive field of the neuron. Then we review some encoding models based on artificial neural networks (ANNs) that use recent state-of-the-art machine learning techniques to address complex natural scenes. For decoding, however, one has to rely on statistical and machine learning models aiming for reconstruction of visual scenes from neuronal spikes. We review some of these decoders with an emphasis on how they can be used for retinal neuroprosthesis to get a better performance for both static images and dynamical videos. 

In Sec.~\ref{net}, sampling-based modeling approach is discussed, where we give an overview of the retinal circuitry in which visual computation can be implemented by probabilistic graph models and spiking neuronal networks, such that different functional networks can conduct visual computations observed in the retina. We first introduce the basis of neural computation with spikes. Some modeling frameworks about neuronal spikes and spiking neural networks (SNNs) are discussed with a sampling perspective. We then propose that studying of the retinal computation should go beyond the classical description of dynamics of neurons and neural networks by taking into account probabilistic inference. We review some of the recent results about how to implement probabilistic inference with SNNs. Traditionally, these approaches are applied to theoretical studies of the visual cortex. Here we demonstrate that how one can use these similar computational approaches for the retinal computation. 

Finally, Sec.\ref{summary} concludes the paper with discussion for some possible research directions in the future.

\section{ Visual computation in neuronal circuit of the retina }\label{retina}

\begin{figure*}[t]
	\begin{center}
		\includegraphics[width=1.5\columnwidth]{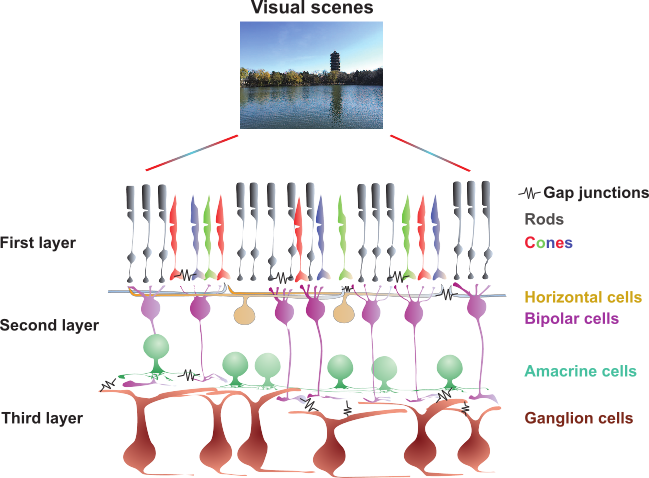}
	\end{center}
	\caption{Illustration of the retinal neuronal circuit. Visual scenes are converted by photoreceptors at the first layer, where rods encode the dim light, and cones encode color. Then the signals, after modulated by horizontal cells, send to bipolar cells at the second layer. The outputs are sent to the third layer consisted of amacrine cells and ganglion cells for further processing. Final signals of the retina are the spikes of ganglion cells transferred to the cortex. Besides chemical synapses between cells, massive gap junctions exist between different and same types of cells, e.g. ganglion-ganglion cells. 
	} 
	\label{fig:retina}
\end{figure*}

Fig.~\ref{fig:retina} shows a typical setup of the retinal neuronal circuit. Roughly, there are three layers of networks consisted of a few types of neurons. Following the information flow of optical visual scenes, photoreceptors convert the light with a wide spectrum of intensities, from dim to bright, and colors, from red, green to blue, into electrical signals that are then modulated by inhibitory horizontal cells. Next, these signals are transferred to excitatory bipolar cells that carry out complex computations. The outputs of bipolar cells are mostly viewed as graded signals, however, the recent evidence suggests that bipolar cells could generate some fast spiking events~\cite{Baden2011}. Then, inhibitory amacrine cells modulate these outputs in different ways to make computations more efficient, specific and diverse~\cite{Jadzinsky2013}. At the final stage of the retina, the signals pass to the ganglion cells for final processing. In the end, ganglion cells send their spikes to the thalamus and cortex for higher cognition. 

Each type of neuron in the retina has a large variation of morphology, for example, it has been suggested that in the mouse retina, there are about 14 types of bipolar cells~\cite{euler2014retinal, franke2017inhibition}, 40 types of amacrine cells~\cite{helmstaedter2013connectomic}, and 30 types of ganglion cells~\cite{Baden2016}. Besides neurons, one unique feature of any neuronal circuitry is the connections between neurons. Typically, connections between neurons in the retina are formed by various types of chemical synapses. However, there are a massive number of electrical synaptic connections, or gap junctions, between different types of cells and within the same type of cells~\cite{bloomfield2009diverse, grimes2018parallel, Brien2018plasticity, rivlin2018flexible}. It remains unclear what is the functional role of these gap junctions~\cite{bloomfield2009diverse}. We hypothesize that gap junctions have a functional role of recurrent connections to enhance visual computation in the retina, which will be discussed in later sections.

In the field of the retinal research, most of the studies are based on the traditional view that neurons in the retina have static receptive fields that are considered as spatiotemporal filters to extract local features from visual scenes. We also know that the retina has many levels of complexities in the information processing, from photoreceptors, bipolar cells, to ganglion cells. In addition, the functional role of modulation of inhibitory horizontal and amacrine cells are still unclear~\cite{Jadzinsky2013, demb2015functional}. Perhaps, the only relative well-understood example is the computation of direction selectivity in the retina~\cite{briggman2011wiring, wei2011organization, zhang2017establishing,mauss2017visual}.

The retinal ganglion cells are the only output of the retina, but their activities are tightly coupled and highly interactive with the rest of the retina. These interactions not only make the retinal circuitry complicated in its structure, but also make the underlying computation much richer for visual processing. Therefore, the retina should be considered smarter than what scientists believed~\cite{Gollisch_2010}. These observations lead us to rethink the functional and structural properties of the retina. Given such a complexity of neurons and neuronal circuits in the retina, we propose that the computations of visual scenes carried by the retina need to go beyond the view that the retina is just like a feedforward network making the information go through. Like the cortical cortex, the retina also has lateral inhibition and recurrent connections (e.g. gap junctions), which make the retina inherit various motifs of neural networks for specific computation of extracting different features of visual scenes, just like the visual processing occurred in the visual cortex~\cite{navlakha2017network,douglas2004neuronal,braganza2018circuit}. 

It should be noted that compared to the visual cortex, the detailed understanding of computation and function of the retina for visual processing has just emerged in recent several decades. Nowadays, the retinal computation of visual scenes by its neurons and neuronal circuits is also refined at many different levels, for details, see recent reviews on neuroscience advancements on the retina~\cite{Jadzinsky2013, euler2014retinal, bloomfield2009diverse, grimes2018parallel, Brien2018plasticity, rivlin2018flexible, demb2015functional, Gollisch_2010}.

\section{Computational framework for the retina}\label{computing}

\begin{figure*}[t]
	\begin{center}
		\includegraphics[width=0.7\textwidth]{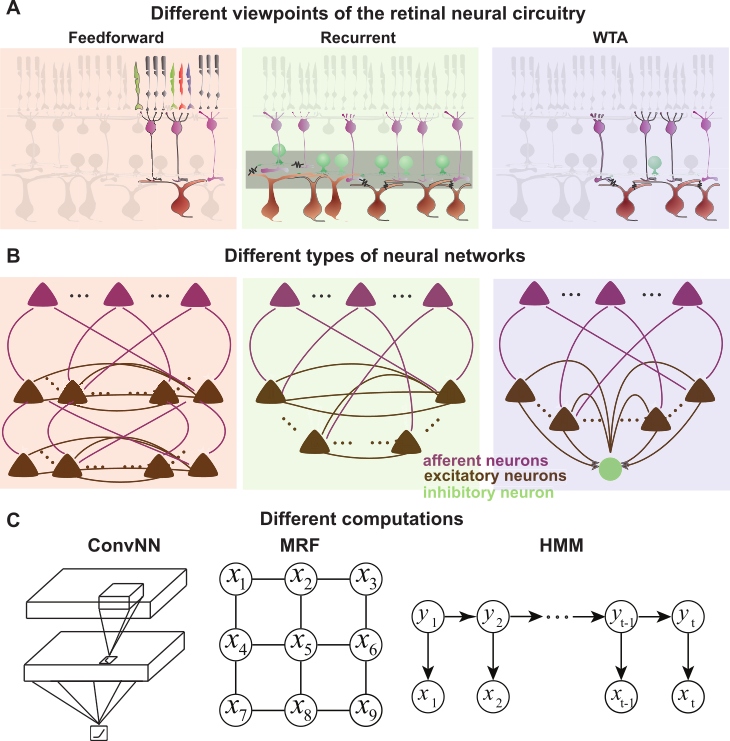}
	\end{center}
	\caption{ Illustration of different computational network motifs. (A) Part of retinal circuity shows different network motif as feedforward, recurrent and winner-take-all (WTA) subnetwork. (B) Abstract representation of different types of neural networks used by modeling, where stimulus is first represented by the activities of afferent neurons, and then feed into a network of excitatory and/or inhibitory neurons for computation. Shadowed networks indicate the same motifs. (C) Abstract computation specifically used by certain typical ANNs, such as convolutional neural networks (ConvNNs), Markov random fields (MRFs) and hidden Markov models (HMMs). Note ANNs can use one or mixed computational network motifs shown in (B). } 
	\label{fig:3net}
\end{figure*}

Different pieces of neuroscience experimental evidence from the retinal circuit seems to be hard to unify from the viewpoint of biology~\cite{werblin2011retinal}. Here we instead hypothesize that one has to study the computation carried out by the retinal circuit with a combination of diverse neural network structure motifs. Such an as-yet-to-emerge computational framework could benefit our understanding of visual computation by utilizing emerged machine learning techniques in recent years~\cite{Lecun2015Deep}. 
When looking at the complete overview of the retinal neuronal circuitry as in Fig.~\ref{fig:retina}, it seems rather complicated. After extracting some features of network structures, there are some simple network motifs emerged. Here we only focus on three types of network structures: feedforward, recurrent and winner-take-all networks as illustrated in Fig.~\ref{fig:3net}, and hypothesize that they play different functional roles in visual computation of the retina. However, the retina is more than a hybrid of these three network motifs, but consists of multiple types of networks to form a hypercircuit~\cite{werblin2011retinal}, where more computational features can be extracted with the advancement of experimental and computational techniques. Such a hypercircuit view provides the biological basis for a potential unified framework of the retinal computation, although how these different networks work together more efficiently for visual computation is still an open question.

\subsection{Feedforward network} 
Feedforward network is the most classical view of the retina as the direction of the visual information flow as in Fig.~\ref{fig:3net}(A-B). Feedforward information flow of the light goes through the retina by three major types of cells, photoreceptors, bipolar cells and ganglion cells. The other two types of inhibitory cells play a modulation role, which has been ignored simply in this viewpoint. The biological basis of this view can be seen from the fovea where excitatory cells play a major role, but inhibitions are little~\cite{sinha2017cellular}. In the fovea, there is a direct cascade processing from photoreceptors, to bipolar cells, and then ganglion cells as outputs.

The advantage of feedforward network has been demonstrated by the advancement of ANNs in recent years, in particular, some breakthroughs have been made in the framework of deep convolutional neural networks (CNNs)~\cite{Lecun2015Deep}. A simple CNN with three layers as in the retina is shown in Fig.~\ref{fig:3net}(C), where a convolutional filter plays a role of the receptive field of retinal cell. A cascade processing of visual inputs is computed by the receptive field of each individual neuron in the retina. The pooling of computation from the previous layer passes to a neuron in the next layer. Recent studies highlight the similarity between the structure of CNNs and retinal neural circuitry~\cite{yan2017revealing, maheswaranathan2018deep}, which will be discussed in later sections. 

\subsection{Recurrent network}
Recurrent network has been seen as a major type in the cortical cortex. The dynamics of recurrent network~\cite{Brunel2000a, Liu2009a, liu2011learning}, together with the diversity of synaptic dynamics and plasticities~\cite{zampini2016mechanisms,Bi2001}, are important for understanding the brain's function. 
Here we hypothesize that recurrent connections are also important for the retina. The formation of recurrent connections in the retina is mainly produced by a massive number of gap junctions as in Fig.~\ref{fig:3net}(A). Unlike chemical synapses, gap junctions are bidirectional or symmetric. Within and between all types of cells in the retina, gap junctions are used to form short connections between neighboring cells. However, the functional role of these gap junctions remains unclear~\cite{bloomfield2009diverse}.

From the computational viewpoint, recurrent connections formed by gap junctions make the retinal circuit like a probabilistic graphical model (PGM) of undirected Markov random field (MRF) as in Fig.~\ref{fig:3net}(B-C). PGM provides a powerful formalism for multivariate statistical modeling by combining graph theory and probability theory~\cite{koller2009probabilistic}. It has been widely used in computer vision and computational neuroscience. In contrast to MRF, there is another type of PGM that is mainly referred to Bayesian network, in which the connections have a direction between nodes. Fig.~\ref{fig:3net}(C) shows one type of Bayesian network, termed hidden Markov model (HMM). In recent years, much effort has been dedicated to implement these PGMs by SNNs, which setup an insightful connection between artificial machine computation by PGMs and neural computation observed in the brain, as well as visual computation in the retina. 

\subsection{Winner-take-all network}
Finally, we hypothesize that the retinal circuit has a computational network unit as winner-take-all (WTA) motif. In the cortical cortex, WTA circuit has  been suggested as a powerful computational network motif to implement normalization~\cite{carandini2012normalization}, visual attention~\cite{itti1998model}, classification~\cite{roy2017online}, and others~\cite{kappel2014stdp}. 

In the retina, there are two types of inhibitory neurons sitting in the first two layers. Horizontal cells target photoreceptors and relay the light information to bipolar cells. Amacrine cells modulate the signals between bipolar cell terminals and ganglion cell dendrites. In both types of cells, there are some specific subtypes that are wide-filed or polyaxonal such that they spread action potentials over a long distance greater than 1 mm~\cite{werblin2011retinal}. From the computational viewpoint, this hypercircuit feature of the retina plays a similar functional role as a WTA network motif. The recent study shows that MRF can be implemented by a network of WTA circuit, which suggest that the WTA could be the minimal unit of probabilistic inference for visual computation~\cite{yu2018winner}. 

\subsection{Rich computation with network motifs}
Above we briefly reviewed the retinal circuitry and pointed out three basic neural network motifs that play a role as units for complex computations conducted in the retina. However, there are more different network motifs suggested in cortical microcircuits~\cite{braganza2018circuit}, and these motifs are also suggested to involved in the retinal computation to form the retinal hypecircuit~\cite{werblin2011retinal}. Such a hypercircuit view of the retina makes most of the methods, which are developed for studying visual processing in the cortex, transfer to investigate the retinal computation that embed rich dynamics beyond the traditional view of the retina~\cite{Gollisch_2010}. In particular, quite a few visual functions have been found to be implemented by some certain types of network mechanisms in the retina, see Ref.~\cite{Gollisch_2010} for a detailed discussion.

Recent computational advancements in the field of ANN make many breakthroughs on visual tasks. For instance, deep CNN is a hierarchical network modeling of visual computation from the retina to inferior temporal part of the cortex~\cite{Yamins2016Using}. These feature-based models take advantages of the receptive field to capture visual features. However, CNN models suffer a few disadvantages for visual computation, for instance, the architecture of CNN is largely lacking design principles, which may be enhanced by biological neural networks in the brain, including the retina~\cite{Marblestone_2016}.

On the other hands, it has been suggested that one needs a hierarchical Bayesian inference framework to understand visual computation~\cite{lee2003hierarchical}. In such a sampling-based modeling approach, statistical computation of visual scenes can be formulated by various types of probabilistic models, where different types of network motifs can implement certain computations~\cite{maass2016searching}. Not only in higher part of the cortex, but also in the visual cortex, there are numerous computational techniques in Bayesian models suitable for visual processing of the retina~\cite{lee2003hierarchical}.

However, these two approaches are not completely separate, and in fact, there are more close interactions between them~\cite{Marblestone_2016}. 
We will explain these ideas by using the retina as a model system in below sections: feature-based approach will be discussed in Sec.\ref{coding}, and sampling-based approach will be discussed in  Sec.\ref{net}. 

\section{Encoding and decoding models of the retina}\label{coding}
Neural coding is one of the central questions for systems neuroscience~\cite{Knill2004,Simoncelli2001Natural, wu2006complete}. In particular, for visual coding, it is to understand how visual scenes are represented by neuronal spiking activities, and in turn, how to decode neuronal spiking activities to represent the given visual information. The retina serves as a useful system to study these questions.

\begin{figure*}[t]
	\begin{center}
		\includegraphics[width=14cm]{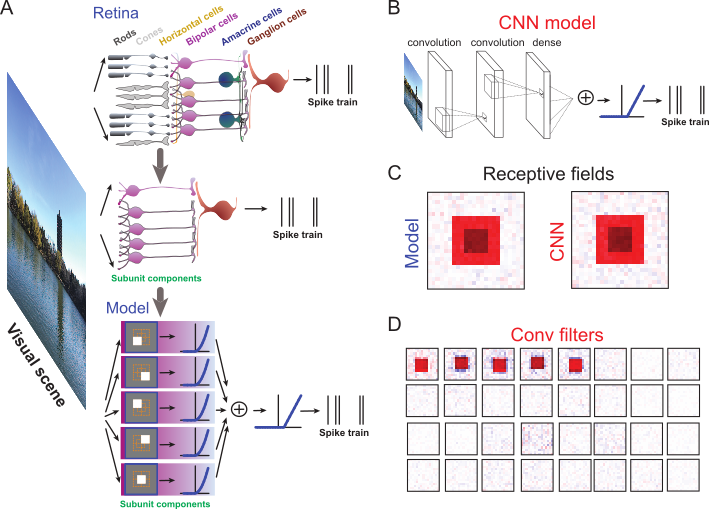}
	\end{center}
	\caption{ Encoding visual scenes by simplified biophysical model with CNN approach. (A) Simplification of a retinal circuitry to a biophysical model. (Top) Feedforward network represented as part of retina circuitry receiving incoming visual scenes and sending out spike trains from ganglion cells. (Middle) Minimal network with one ganglion cell and five bipolar cells. (Bottom) Biophysical model with five subunits representing five bipolar cells, where each has a linear filter as the receptive field, and a nonlinearity. The outputs of five subunits are pooled and rectified by another output nonlinearity. The final output can be sampled to give a spike train. (B) Representative CNN model trained with images as input and spikes as output. Here there are two convolutional layers and one dense layer. (C) After training, the CNN model shows the same receptive field as the biophysical model for modeled ganglion cell. (D) Convolutional filters after training resemble the receptive fields used by the biophysical model of bipolar cells in (A). (A) is adapted from Ref.~\cite{jia2018neural}. (B-D) are adapted from Ref.~\cite{yan2017revealing, yan2018revealing}.  } 
	\label{fig:cn}
\end{figure*}

\subsection{Biophysical encoding model}
For understanding the encoding principles of the retina, quite a few models are developed based on biophysical properties of neurons and neuronal circuits in the retina, which have been reviewed recently~\cite{meyer2017models}. Here we briefly review some approaches.

The starting point for looking at retinal neuronal computation is to find the receptive fields (RFs) of neurons. The classical approach to mapping the neuronal RF is to patch a single cell and then vary the size of a light spot to obtain the RF structure as a difference-of-Gaussian filter with center excitation and surround inhibition. Later on, a systematic experimental method was developed by using multielectrode array to record a population of retinal ganglion cells, in which one can manipulate light stimulation with various types of optical images, including simple bars, spots, gratings, white noise, and complex well-controlled images and movies. In particular, one can analyze the spike trains of individual neurons when recording a large population at one time with white noise stimulus. A simple reverse correlation technique, termed spike-triggered average (STA)~\cite{Chichilnisky2001a}, can obtain the RF of every recorded ganglion cell. An extension of STA to covariance analysis, termed spike-triggered covariance, serves as a powerful tool for analyzing the 2nd order dynamics of the retinal neurons~\cite{Schwartz2006,Liu_2015}.

With the receptive field mapped from each neuron, a simple and useful analysis is based on a linear-nonlinear (LN) model to simulate the cascade processing of light information. 
There are two stages in the LN model~\cite{sahani2003linear,machens2004linearity}. The first stage is a linear spatiotemporal filter encoding the way of integrating inputs, which represents the sensitive area of the cell, i.e., 
the characteristic of the receptive field. The second stage is a nonlinear transformation to convert the output of the linear filter to a firing rate. Both properties of the LN model can be easily estimated from the spikes with white noise stimulus~\cite{Liu_2015}. Otherwise, for complicated stimulus signals rather than white noise, one has to use other methods, such as maximum likelihood estimation~\cite{sahani2003linear} and maximally informative dimensions~\cite{Sharpee2004}, to estimate the model components when there are enough data.

Until now, quite a few models are developed to refine the building blocks of LN model to advance the model to be more powerful, such as linear-nonlinear Poisson model~\cite{Schwartz2006}, where after nonlinear operation, a Poisson process is used to determine whether a spike would be generated; 
generalized linear model~\cite{Pillow2008Spatio}, where several more components are included, such as a spike history filter for adaptation, and a coupling filter for influence of nearby neurons. Recently, the models with a few components of subunits to mimic upstream nonlinear components are emphasized, such as nonlinear input model~\cite{McFarland2013}, where a few upstream nonlinear filters are included with the assumption that the input of the neuron is correlated; 
spike-triggered covariance model~\cite{paninski2003convergence, Rust2005, Liu_2015}, where covariance of spike-triggered ensemble is analyzed with eigenvector analysis to obtain a sequences of filters as a combination of some parts of receptive field; 
2-layer linear-nonlinear network model~\cite{maheswaranathan2018inferring}, where a cascade process is implemented by 2-layer LN models; 
spike-triggered non-negative matrix factorization (STNMF) model~\cite{Liu2017}, where the orthogonality constraint used in spike-triggered covariance is relaxed to obtain a sets of non-orthogonal subunits shown as the bipolar cells in the retina.
It has been further shown that STNMF can recover various biophysical properties of upstream bipolar cells, including spatial receptive fields, temporal filters, transferring nonlinearities, synaptic connection weights from bipolar cells to ganglion cell. In addition, a subset of spikes contributed by each bipolar cell can also be teased apart from the whole spike train of one ganglion cell~\cite{jia2018neural}.

\begin{figure*}[t]
	\begin{center}
		\includegraphics[width=\textwidth]{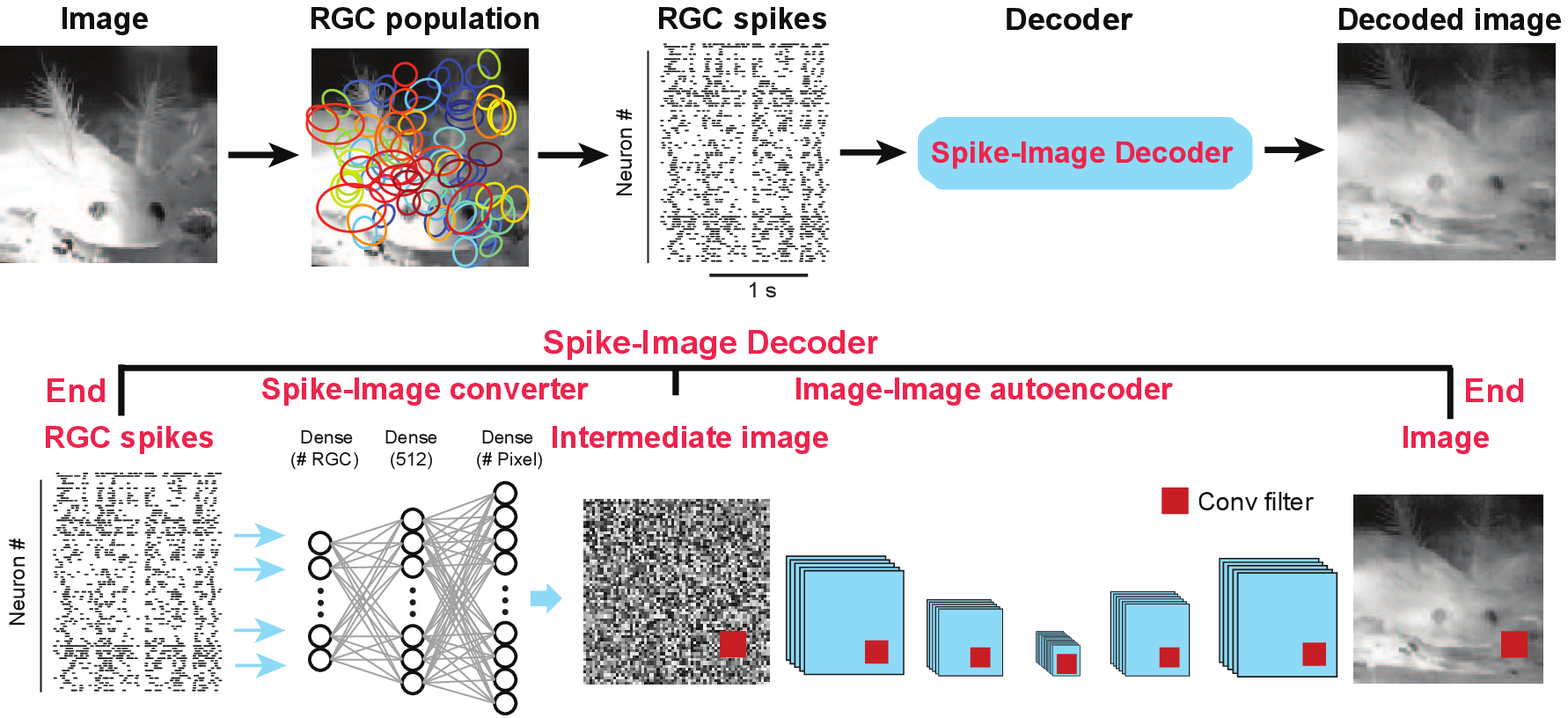}
	\end{center}
	\caption{ 
    Decoding visual scenes from neuronal spikes. (Top) Workflow of decoding visual scenes. Here a salamander swimming video was presented to a salamander retina to get a population of ganglion cells fired with a sequence of spikes. A population of spike trains is used to train a decoder, spike-image decoder, to reconstruct the same video. Receptive fields of ganglion cells are mapped onto the image. Each colored circle is an outline of receptive field. (Bottom) Spike-image decoder is an end-to-end decoder with two stages: spike-image converter used to map a population of spikes to a pixel-level preliminary image, and image-image autoencoder for mapping every pixel to the target pixels in the desired images. Note that there is no unique architecture of spike-image decoder, and the state-of-art model could be adopted and optimized. Exact preliminary images depend on the loss functions used for training. Details of the decoding process can be found in Ref.~\cite{zhang2019reconstruction} and online (see Sec.~\ref{data}). 
    } 
	\label{fig:decoder}
\end{figure*}

\subsection{ANN-based encoding model}
In recent years, ANNs, such as deep CNNs and probabilistic graph models, make some breakthroughs for numerous practical tasks related to system identification of visual information~\cite{Lecun2015Deep}. For instance, with a large set of visual images collected and well-labeled by specific tags, ANNs can outperform the human-level performance for object recognition and classification~\cite{Lecun2015Deep}. Various techniques have been developed to visualize the features of images learned by CNN. However, 
end-to-end learning of complex natural images makes CNN not very interpretable for the underlying network structure components~\cite{Zeiler2014Visualizing,Smith_2016}.

Inspired by experimental observation in neuroscience~\cite{Hassabis2017,Marblestone_2016}, a typical deep CNN has a hierarchical architecture with many layers~\cite{Lecun2010Convolutional}. Out of these layers, there are some layers having a bank of convolutional filters, such that each convolutional filter is served as a feature detector to extract important property of images~\cite{Simonyan2014Very,Krizhevsky2012ImageNet}. Therefore, after training with a large set of images, these convolutional filters play a functional role as neurons in our retina and other visual systems to encode complex statistical properties of natural images~\cite{Simoncelli2001Natural}. The shapes of these filters are sparse and localized, and like receptive fields of visual neurons.  

Therefore, it is not trivial to use similar the ANN-based approache to investigate the central question of neuronal coding in neuroscience~\cite{Kriegeskorte2015,Yamins2016Using}. In particular, for visual coding, it has been widely accepted that the ventral visual pathway in the brain is a path starting from the retina, lateral geniculate nucleus, then layered visual cortex to reach inferior temporal part of the cortex. This visual pathway has been suggested as the ``what pathway'' for recognition and identification of visual objects. When CNN is used to model experimental neuroscience data recorded in neurons of inferior temporal cortex in monkeys, neuronal response can be predicted very well~\cite{Yamins2013Hierarchical,Yamins_2014,Khaligh-Razavi2014, Yamins2016Using}. 
Therefore, it is possible to relate the biological underpinnings of visual processing in the brain with those network structure components used in CNN. However, it is not straightforward to interpret this relationship since the pathway from the retina to inferior temporal cortex is complicated~\cite{Yamins2016Using}. One possible and easier way is to use CNN to model the early visual system of the brain, in particular, the retina as introduced above, in which neuronal organization is relatively simple. 

Indeed, a few studies take this approach by using CNNs and their variations to earlier visual systems in the brain, such as the retina~\cite{yan2017revealing,maheswaranathan2018deep,Batty2017, vance2018bioinspired, yan2018revealing}, V1~\cite{vintch2015convolutional, Antolik2016, kindel2017using,cadena2017deep,Klindt2017}, and V2~\cite{rowekamp2017cross}. Most of these studies are driven by the goal that the better performance of neural response can be archived by using either feedforward and recurrent neural networks (or both). These new approaches increase the complexity level of system identification, compared to conventional linear/nonlinear models~\cite{McFarland2013}. Some of these studies also try to look into the detail of network component after learning to see if and how it is comparable to the biological structure of neuronal networks~\cite{yan2017revealing, maheswaranathan2018deep, Klindt2017}. 

Fig.~\ref{fig:cn} shows a typical setup of CNN modeling approach for the retina. To understand the fine structure of the receptive field in the retinal circuit, this is important to understand the filters leaned by CNNs. In contrast to the studies where a population of retinal ganglion cells are used~\cite{maheswaranathan2018deep,Klindt2017, mcintosh2016deep}, one can simplify the model from a complicated retinal circuit to a simple network model as in Fig.~\ref{fig:cn}(A), which makes the modeling easier to refine the structure components at the single cell level of the retina. Indeed, it is found that CNNs can learn their internal structure components to match the biological neurons of the retina~\cite{maheswaranathan2018deep, yan2018revealing}, as illustrated in Fig.~\ref{fig:cn} (D). 

Given that the retina has a relatively clear and simple circuit, and the eyes have (almost) no feedback connection from the cortical cortex, it is a suitable model system as a feedforward neural network, similar to the principle of CNN. Certainly, the contribution from the inhibitory neurons, such as horizontal cells and amacrine cells, play a role for the function of the retina. In this sense, the potential neural networks with lateral inhibition and/or recurrent units are desirable~\cite{Batty2017,mcintosh2016deep}.

\subsection{Decoding visual scenes from retinal spikes}
From the viewpoint of retinal neuroprosthesis, an ideal encoder model is able to deliver precise stimulation to electrodes with given visual scenes. For this, one has to close the loop to find an ideal decoder model that can readout and reconstruct stimulus of visual scenes from neuronal responses. 

Reconstruction of visual scenes has been studied over many years. The neuronal signals of interest can be fMRI human brain activities~\cite{thirion2006inverse,naselaris2009bayesian,nishimoto2011reconstructing,wen2017neural}, neuronal spikes in the retina~\cite{marre2015high,Parthasarathy2017,botella2018nonlinear,gollisch2008rapid} and lateral geniculate nucleus~\cite{Stanley1999}, neuronal calcium imaging data in V1~\cite{garasto2018visual}. However, the decoding performance of current methods is rather low for natural scenes, either static natural images or dynamical movies. One particularly interesting example of the movies reconstructed from fMRI data can be found from Ref.~\cite{nishimoto2011reconstructing}.

For the retinal neuroprosthesis, one would expect to decode visual scenes by using spiking responses of a population of ganglion cells. Decoding of visual scenes is possible when there are enough retinal ganglion cells available as shown in a recent study with simulated retinal ganglion cells~\cite{Parthasarathy2017}. However, it is unclear whether one can use experimental data to achieve this aim. One can name this decoding approach as a spike-image decoder that performs an end-to-end training process from neuronal spikes to visual scenes. 

Recently, we developed such a decoder with a model of deep learning neural network that can achieve a much better resolution than previous studies for reconstructing natural visual scenes, including both static images and dynamic videos, from spike trains of a population of retinal ganglion cells recorded simultaneously~\cite{zhang2019reconstruction}. 

The workflow of the spike-image decoder is illustrated in Fig.~\ref{fig:decoder}. With a setup of multi-electrode array, a large population of the retinal ganglion cells can be recorded simultaneously, and their spikes can be extracted. Then, a spike-image converter is used to map spikes of every ganglion cell to images at the pixel level. After that, one can apply autoencoder deep learning neural network to transfer/enhance spike-based images to original stimulus images. Essentially, this approach has two stages with one as spike-image converter and the other as image-image autoencoder. Most of the previous studies focused on the first stage, which is the traditional decoder to be optimized by some statistical models and/or ANN-based models in either linear or nonlinear fashion~\cite{thirion2006inverse, naselaris2009bayesian, nishimoto2011reconstructing, wen2017neural, marre2015high,Parthasarathy2017, botella2018nonlinear, gollisch2008rapid, Stanley1999}. A recent study trained a separate CNN autoencoder as the second stage to enhance the quality of images~\cite{Parthasarathy2017}. Instead, we found a better quality can be achieved by the end-to-end training process with both stages of spike-to-image converter and image-to-image autoencoder together. However, the detailed architecture of networks used in these two stages could be optimized to an even better quality with other possible deep learning neural networks.



\begin{figure*}[t]
	\begin{center}
		\includegraphics[width=13cm]{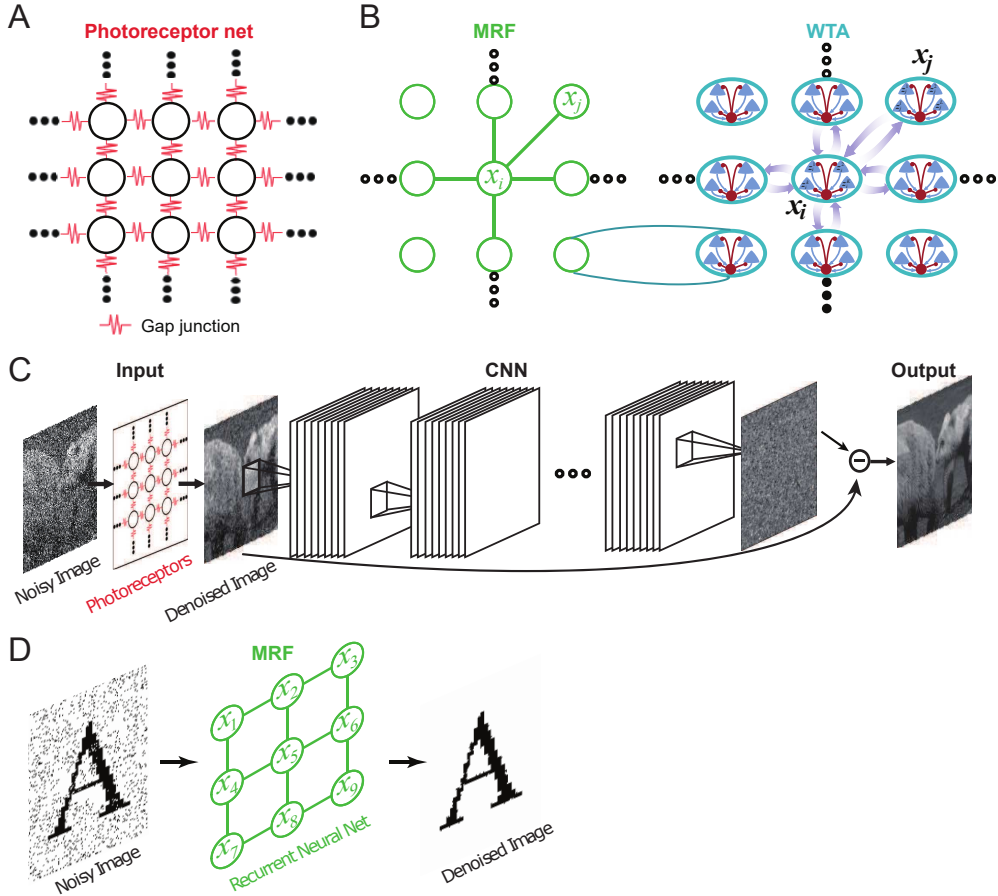}
	\end{center}
	\caption{ Implementation of noise reduction computation with the retinal photoreceptors, probabilistic graph model and spiking neural network.  (A) A network of rod photoreceptors connected by gap junctions. (B) A graph of Markov random field (MRF) represented by a network of spiking neurons with subnetworks as winner-take-all (WTA) circuits. Each variable of MRF is represented by one WTA neural network. (C) Noisy images can be denoised by photoreceptor network, then enhanced by CNN. (D) Noisy images can be denoised by MRF implemented by a recurrent spiking neuron network without enhancement. (A) and (C) are adapted from Ref.~\cite{yue2018simple}. (B) is adapted from Ref.~\cite{yu2018winner}. (D) is adapted from Ref.~\cite{zheng2019probabilistic}. } 
	\label{fig:pr}
\end{figure*}

\section{Modeling the retina with SNNs and PGMs }\label{net}
SNNs are thought as the third generation of ANN models, which use neuronal spikes for computation as in the brain~\cite{maass1997networks}. 
Together with neuronal and synaptic states, the importance of spike timing is also considered in SNNs. It has been proved that SNNs are computationally more powerful than other ANNs with the same number of neurons~\cite{maass1997networks}. In recent years, SNNs have been widely studied in a number of research areas~\cite{ghosh2009spiking, paugam2012computing,jonke2016solving}. In particular, recent studies show that SNNs can be combined with a deep architecture of multiple layers to obtain the similar or better performance as ANNs~\cite{tavanaei2018deep,hu2018spiking, qi2018jointly, xu2018csnn, xiao2018spike}. In addition, the spiking feature of SNNs is particularly important for the next generation of neuromorphic computer chips~\cite{esser2016convolutional,davies2018loihi}.  

The computational capability of a single neuron is limited. However, when a population of neurons are connected together to form a network, their computational ability can be greatly expanded. In terms of the language of graphs~\cite{maass1996lower}, a SNN can be denoted as a graph $G=(V, E)$, of which $V$ represents the set of neurons and $E \subset V \times V$ represents the set of synapses. 
Given this equivalence between graphs and neural networks, a different approach termed probabilistic graphical models (PGMs), has also been intensively studied over recent years. The idea of PGM is in that, traditionally, both ANNs and SNNs are doing modeling as a deterministic dynamical system, which has been demonstrated by the classical Hodgkin-–Huxley model~\cite{hodgkin1952quantitative}. However, the computational principles used in the brain seem to go beyond this viewpoint \cite{maass2016searching}.  

There is an increasing volume of neuroscience evidence that humans and monkeys (other animals as well) can represent probabilities and implement probabilistic computation~\cite{meyniel2015confidence, ernst2002humans,kording2004bayesian}, and the viewpoint of the probabilistic brain is increasingly recognized~\cite{Pouget2013Probabilistic}. Therefore, the network of spiking neurons has been used to implement probabilistic inference at the neural circuit level~\cite{Pouget2013Probabilistic}. The combination of SNNs with probabilistic computation shows an increasing research interest for both understanding the principles of brain computation and solving practical problems with these brain-inspired principles. 

Traditionally, probabilistic inference studied in the framework of PGM is a combination model of probability theory and graph theory.
The core idea of PGMs is taking advantage of a graph to represent the joint distribution among a set of variables, of which each node corresponds to a variable and each edge corresponds to a direct probabilistic interaction between two variables. With the benefit of a graph structure, a complex distribution over a high-dimensional space can be factorized into a product of low-dimensional local potential functions. PGMs can be divided into directed graphical models, such as Bayesian network, and undirected graphical models, such as Markov random fields. 
Bayesian networks can represent causality between variables, so they are often used to model the process of cognition and perception. 
While Markov random fields can represent a joint distribution by a product of local potential functions.

Implementing PGMs by SNNs is to explain how neuronal spikes can implement probabilistic inference. Inference in SNNs includes two main questions. 1.) Probabilistic coding: how neural activities of a single cell or a population of cells (like membrane potential and spikes) encode probability distribution. 2.) Probabilistic inference: how the dynamics of a network of spiking neurons approximate the inference with probabilistic coding. 

Obviously, probabilistic coding is the precondition of probabilistic inference. According to the way of expressing probability, probabilistic codes can be divided into three basic types: 1) those that encode the probability of each variable in each state, such as probability code~\cite{anastasio2000using}, log-probability code~\cite{rao2004bayesian,rao2004hierarchical}, and log-likelihood ratio code~\cite{gold2002banburismus,deneve2008bayesian}, 2) those that encode the parameters of a distribution, such as probabilistic population code that takes advantage of neural variability~\cite{milton2011noisy,fisher2004small,ringach2009spontaneous}, that is, neural activities in response to a constant stimulus have a large variability, which suggests that population activities of neurons can encode distributions automatically, 3) those that consider neural activities as sampling from a distribution~\cite{hoyer2003interpreting, fiser2010statistically}, which has been suggested by numerous experiments~\cite{bonawitz2014probabilistic,denison2013rational,gopnik2012reconstructing,bonawitz2012children}. 

According to these coding principles, there are different ways to implement inference with a network of neurons: 1) implementing inference with neural dynamics that has equations similar to the inference equations of some probabilistic graphical models over the time course~\cite{rao2004bayesian, rao2004hierarchical, deneve2008bayesian, beck2007exact,yu2017neural,  ott2006neurodynamics}. This approach is mainly suitable for small-scale SNNs; 2) inference with neural variational approximations that is suitable to describe the dynamics of a large-scale SNN directly \cite{yu2018winner,lee2003hierarchical, george2006belief,george2009towards, steimer2009belief, litvak2009cortical,friston2009free,friston2010free,friston2017graphical,friston2017active}; 3) inference with probabilistic population coding and some neural plausible operations, including summation, multiplication, linear combination and normalization~\cite{ma2006bayesian, beck2008probabilistic, ma2013towards,beck2011marginalization,yu2016sampling}; 4) inference with neural sampling over time where the noise, such as stochastic neural response found in experimental observations~\cite{yang2007probabilistic,gold2007neural}, is the key for neural sampling and inference~\cite{pecevski2011probabilistic, buesing2011neural, habenschuss2013stochastic, legenstein2014ensembles, maass2014noise}. Similarly, one can do the sampling by using a large number of neurons to sample from a distribution at the same time~\cite{ yu2016sampling, shi2009neural, shi2010exemplar, yu2018implementation}, as it is found that the states of neurons in some areas of the brain follow special distributions~\cite{coppola1998unequal, furmanski2000oblique}. 

Although the above studies are mostly conducted in an abstract way for neural computation of the cortex, including the visual cortex. Here we suggest that these computational techniques can be transferred to study the retinal computation. Fig.~\ref{fig:pr} shows some examples in the retina where there is a similarity at the network level between a network of photoreceptors connected by gap junctions (Fig.~\ref{fig:pr}(A)), a Markov random field model (Fig.~\ref{fig:pr}(B)), and an implementation of MRF by a network of spiking neurons consisted of clusters of winner-take-all microcircuits (Fig.~\ref{fig:pr}(B)). As illustrated in Fig.~\ref{fig:3net}, massive gap junctions play a functional role of recurrent connections between retinal neurons. A recent study shows that a network of rod photoreceptors with gap junctions can denoise images that can be further enhanced by an additional CNN as in Fig.~\ref{fig:pr}(C). It is found that this CNN with photoreceptors included, comparing to other traditional CNNs, can achieve state-of-the-art performance for denoising~\cite{yue2018simple}. Similarly, PGM has also been used to denoise images~\cite{bishop2006pattern}. Recently, it is shown that PGMs can be implemented by SNNs for various types of computations~\cite{ yu2018winner, yu2018implementation, yu2018unification, Yu_2016, guo2017hierarchical, yu2018emergent}, thus, when using SNNs for denoising, similar performance can be achieved~\cite{zheng2019probabilistic}, as illustrated in Fig.~\ref{fig:pr}(D).

Probabilistic graphical models have been intensively studied and used for visual coding, but mostly for the cortical process~\cite{lee2003hierarchical}. Here these results suggest that one can study visual computation in the retina by combing several approaches into a systematical framework, including classical PGMs, nontrivial retinal circuit structures, in particular, gap junctions, and recent efforts about the implementation of PGMs by SNNs. Future work is needed to study this framework with more inspirations from the rich network structure of the retina, including recurrent neural network, winner-take-all circuit, and feedforward neural network, even other ubiquitous motifs of cortical microcircuits~\cite{braganza2018circuit}.

\section*{ Discussion }
\label{summary}

Neuroprosthesis is a promising medical device within the framework of precision medicine. With directly talking to the brain of each individual patient, neuroprosthesis needs to be advanced with better computational algorithms for neuronal signal, besides better hardware designs. For the computational capability of the retinal neuroprosthesis, the major difficulty is in that one has to track the complexity of spatiotemporal visual scenes. 

In contrast to other neuroprostheses, where incoming signals are in a low dimensional space, such as moving trajectory of body arms/legs in 3D space, or auditory signal in a 1D frequency space, our visual scenes are more complex with information in a spatiotemporal fashion. Recent advancements of computer vision make some breakthroughs for analyzing these complex natural scenes, which make a wave of artificial intelligence up to a high attitude than ever before. 

On the other hand, with experimental advancements in neuroscience, one can collect a large population of neurons simultaneously. In particular, in the retina, a population of spike trains from hundreds of retinal ganglion cells can be obtained with well-controlled visual scenes, such as images and movies~\cite{Onken_2016}. The newest technique can record several thousands of neurons simultaneously~\cite{portelli2016rank, maccione2014following,hilgen2017pan}. This opens the gate for studying the encoding and decoding of visual scenes by using enough spikes to achieve a superb resolution.  

Out of the current approaches for retinal neuroprosthesis, the implants with electrodes are the mainstream and have been used in clinical trials. However, there are very limited computational models embedded into the retinal prosthesis~\cite{Nirenberg2012, yue2016retinal, yan2018embedded}. With an encoder embedded, it is possible to process incoming visual scenes to better trigger ganglion cells~\cite{Nirenberg2012, yue2016retinal}. The benefit of decoding models is to justify the spiking patterns produced by the targeted downstream neurons. Ideally, electrical stimulation should be able to close to those desired patterns of retinal neural activity in a prosthesis. To compare the similarity between spiking patterns, the traditional way focuses on how to compute the distance between two spike trains in general~\cite{victor2005spike, rossum2001novel}, and in the context of the retinal prosthesis~\cite{shah2017learning}.
Another way of doing this is to using decoding models for the purpose of better performance of neuroprosthesis~\cite{li2014decoding, Parthasarathy2017, Nirenberg2012}.
Ideally, similar to the other neuroprostheses, where a closed-loop device can be employed to decode neuronal signal to control stimulus, the signal delivered by a retinal prosthesis should be able to reconstruct the original stimuli, i.e., dynamic visual scenes projected into the retina. Thus, one can use a decoding model to reconstruct visual scenes from the spiking patterns of the retinal ganglion cells~\cite{Nirenberg2012, Parthasarathy2017}. Such a direct measure of the precision of spiking patterns with the given decoding model could play a functional role of controlling electrical stimulation patterns generated by the retinal neuroprosthesis, which is the goal of a better and adjustable neuroprosthesis.

Here we only focused on the computational modelling issue of one type of retinal neuroprosthsis with electrodes embedded. Certainly, for retinal neuroprosthsis, as an engineering system, there are many parallel difficult issues, such as advanced materials, power designing, communication efficiency, and other related hardware issues, which have been covered by many well-written reviews~\cite{ghezzi2015retinal, goetz2016electronic, yue2016retinal, shepherd2013visual}. One should note that there are different types of visual implants, including those with light retinal stimulation such as optogenetics and chemical photoswitches, as well as implants in other parts, beyond the retina, of the brain. The computational issues raised in this paper are also relevant to the general visual prosthesis. Besides these artificial visual implants, another line of researches focuses on retinal repair by biological manipulation of stem cells, such as induced pluripotent stem cells~\cite{al2014retinal,borooah2013using,mandai2017autologous}, where understanding the computational mechanisms of biological neurons and neuronal circuits is more relevant to encoding visual scenes. For which, the potential decoding models may need more efforts to include the biological principles found in the retina~\cite{Gollisch_2010}.

Taken together with all these advancements of neuroscience experiment and prosthesis engineering, now it is time to advance our understanding of visual coding by using the retinal spiking data and ANN-based models to get better computational algorithms for improving the performance of retinal neuroprosthesis. Here we reviewed some of the recent progress on developing novel functional artificial intelligence models for visual computation. Feature-based modelling approach, such as deep CNN, has made significant progresses on analysis of complex visual scenes. For some particular visual tasks, these models can outperform the human~\cite{Lecun2015Deep}. However, the efficiency, generalization ability, and adaption or transfer learning between different tasks, of well-trained models are still far from human performance~\cite{Marblestone_2016}. Sampling-based modelling with neuronal spikes emerges as a new approach, which takes advantage of many factors of the neuronal system of the brain~\cite{maass2016searching}, such as noise at the level of single neurons and synapses~\cite{maass2014noise, buesing2011neural, kappel2014stdp}. With the generic benefit of pixel representation of visual scenes, sampling models can be naturally used for various types of visual computations~\cite{bishop2006pattern}. However, the efficiency of learning algorithms of sampling model is still far from the flexibility of the neuron system of the brain~\cite{yu2016camkii}. Nevertheless, these two approaches could be combined by utilizing both advantages of feature and sampling for visual computation. For this, one needs to consider the retina as a neuronal network where visual computation can be done by different functional network structures. The future work is needed to combine various network motifs into a hybrid network, in which different visual information can be extracted, processed, and computed. Such hybrid or hypercircuit networks have been explored only in very recent years so far, in particular, WTA network motif has been shown as a functional module in more complex hypercircuit network model for various types of computations~\cite{kappel2014stdp, yu2018winner, jonke2016solving, rutishauser2018solving}. One expects that there will be more studies align this line in future.

The modelling framework mentioned in this paper is not limited to the application of the retina, but could be used to other visual systems in the brain, and to other artificial visual systems. The main feature of these algorithms is to make use of neural spikes. Advancements of recent artificial intelligence computing align with the development of the next generation of neuromorphic chips and devices, where the new data format is processed as spikes or events~\cite{dong2017spike, bi2018spike, lichtsteiner2008128, li2017cifar10, orchard2015converting}. Therefore, the methods can be applied for neuromorphic visual cameras with spike or event signals as well. One can use these computational retinal models to simulate a population of spikes for encoding and decoding of any given visual scenes, including static natural images, dynamic videos, even real-time videos captured by standard frame-based camera~\cite{zhang2019reconstruction}. Taken neuromorphic hardware and event/spiking computing algorithm together, the next generation of computational vision can develop a better system for artificial vision beyond the purpose of retinal neuroprosthesis. 
Therefore, we believe that rich interactions between artificial intelligence, computer vision, meromorphic computing, neuroscience, bioengineering, and medicine, will be important for advancing our understanding of the brain, and developing the next generation of retinal neuroprosthesis for artificial vision system. The algorithm part of the artificial eye, including encoding and decoding models of natural visual scenes, will be in particular crucial for such a systems-level approach.


\section*{Data availability}
\label{data}
Data presented in Fig.~\ref{fig:decoder} are publicly available online: Retinal experimental data demonstrated are available at dx.doi.org/10.5061/dryad.4ch10. Reconstruction examples are available at https://sites.google.com/site/jiankliu. 




\section*{References}
  \bibliographystyle{elsarticle-num} 
  \bibliography{ref_all_new_nolink2endnote}





\end{document}